\DocumentMetadata{
	lang		= eng-US, 	% default
	pdfstandard	= A-3u,		% A-2b, A-2u, A-3b, A-3u. Check that your figures also meet the standard you choose.
	pdfversion	= 1.7, 		% an older version of PDF, but very widely supported
}

\documentclass[colorlinks,upint,subscriptcorrection,varvw,hyphenate,balance,greek,russian,vietnamese,german]{asmeconf} % pdftex

\hypersetup{%
	pdfauthor={John H. Lienhard},									  % <=== change to YOUR name
	pdftitle={ASME Conference Paper LaTeX Template},                  % <=== change to YOUR pdf file title
	pdfkeywords={ASME conference paper, LaTeX template, BibTeX style},% <=== change to YOUR pdf keywords
	pdfsubject = {Describes the asmeconf LaTeX template},			  % <=== change to YOUR subject
}
\allowdisplaybreaks % from amsmath package, allows multiline equations to break across pages (delete if not wanted)
\begin{document}

% Change these fields to the right content for your conference.
% You can comment these out if for some reason you don't want a header.
% Use title case for the conference name (first letters capitalized), not all capitals

\ConfName{Proceedings of the ASME 2026\linebreak Fluids Engineering Division Summer Meeting }
\ConfAcronym{FEDSM2026}
\ConfDate{July 26--29, 2026} 
\ConfCity{Bellevue, WA}
\PaperNo{FEDSM2026-184520}

% Units of measure (e.g., cm) and other specialty lowercase terms in the title should be 
%   enclosed in \NoCaseChange{...} to maintain lower case type.
%   The rest of the title will automatically be set in all capital letters.
%
%	\title{Place Title Here: Place Subtitle After Colon} 

\title{Laminar-to-Turbulent Transition of Yield-Stress Fluids in Pipe and Channel Flows} % <=== replace with YOUR title
 
%   Put author names into the order you want. Use the same order for affiliations.
%   \affil{#} tags the author's affiliation to the address in \SetAffiliation{#}.
%   No space between last name and \affil{#}, separate names with commas.
%
%	For a sole author or a single affiliation for all authors, {#} may be left empty, i.e. \affil{} and \SetAffiliation{} (but not with [grid] option!)
%
%   \CorrespondingAuthor{email} follows that author's affiliation, no spaces before or after 
%   If multiple corresponding authors, put both email addresses in the same command and place after both authors.
%
%   \JointFirstAuthor, if applicable, follows the affiliation of the relevant authors, no spaces.

\SetAuthors{%
  Shivam Prajapati\affil{1}\affil{2}\CorrespondingAuthor{sprajapati31@gatech.edu},
  Prasoon Suchandra\affil{1}\affil{2},
  Vivek Kumar\affil{1}\affil{2},
  Ardalan Javadi\affil{1},
  Suhas Jain\affil{1},
  Cyrus Aidun\affil{1}\affil{2}
}

%	Note: Luis and Maria are not real people.  Henry and Catherine have been dead for >450 years.

\SetAffiliation{1}{George W. Woodruff School of Mechanical Engineering, Georgia Institute of Technology, Atlanta, GA 30332, USA}
\SetAffiliation{2}{Renewable Bioproducts Institute, Georgia Institute of Technology, Atlanta, GA 30332, USA}

\maketitle

\versionfootnote{Documentation for \texttt{asmeconf.cls}: Version~\versionno, \today.}% <=== Delete before final submission.

\keywords{Herschel--Bulkley fluid, laminar--turbulent transition, direct numerical simulation, yield-stress}

%%%% Abstract should be 200 words or less
\begin{abstract}
We present direct numerical simulations (DNS) of laminar to turbulent transition in Herschel--Bulkley (HB) yield-stress fluids flowing through pipes and rectangular channels. The simulations employ a Herschel--Bulkley formulation that captures the yield-stress-driven plug, its breakdown, and the emergence of near-wall turbulent structures, enabling direct resolution of the transition mechanisms.

The DNS cover a broad range of generalized Reynolds numbers, $Re_{G}=378$ to $5300$, allowing us to resolve plug formation, transition onset, and fully turbulent regimes. In pipe flow, the simulations reproduce the characteristic transition sequence, which includes a strong plug and negligible turbulence at low $Re_{G}$, a sharp rise in turbulence intensity and $u_{\mathrm{rms}}^{\prime +}$ within a narrow transitional window ($Re_{G}\approx 2000$ to $3000$), and wall-dominated turbulence with a weakened core at higher $Re_{G}$. Transition occurs only when local Reynolds stresses exceed the yield stress. The resulting regime boundaries ($Re_{G}<1735$ laminar, $1735<Re_{G}<2920$ transitional, and $Re_{G}>2920$ turbulent) align with trends reported for Carbopol fluids.

DNS of channel flow are performed in a $0.2\,\mathrm{m}$ streamwise domain using a $128\times129\times128$ grid (approximately $4.2$ million cells). The simulations capture the same HB transition physics, including formation of a central plug, growth of near-wall turbulence observed via Q-criterion structures, and progressive strengthening of $u'_{\mathrm{rms}}$ and turbulent kinetic energy peaks with increasing $Re_{G}$. DNS also highlights the strong influence of shear-thinning and yield stress on transitional behavior.

This work provides the first DNS resolving the complete laminar to turbulent transition in HB fluids for both pipe and channel configurations, offering unified insight into plug breakdown, turbulence localization, and the role of yield stress in transition mechanisms. Experimental validation using a 3.6\,m acrylic channel with particle image velocimetry (PIV) is planned to further assess the DNS predictions and quantify geometry-dependent transition thresholds.
\end{abstract}

\begin{nomenclature}

\EntryHeading{Roman letters}
\entry{${u}$}{velocity [m\,s$^{-1}$]}
\entry{$p$}{pressure [Pa]}
\entry{$u_\tau$}{friction velocity [m\,s$^{-1}$]}

\EntryHeading{Greek letters}
\entry{$\rho$}{density [kg\,m$^{-3}$]}
\entry{$\tau_y$}{yield stress [Pa]}
\entry{$\kappa$}{consistency index [Pa\,s$^{n}$]}
\entry{$n$}{power-law index [-]}

\EntryHeading{Dimensionless / statistics}
\entry{$Re_G$}{generalized Reynolds number [-]}
\entry{$Re_\tau$}{friction Reynolds number [-]}
\entry{$u'_{\mathrm{rms}}$}{r.m.s. velocity fluctuation [m\,s$^{-1}$]}
\entry{$I$}{turbulence intensity [$\%$]}

\end{nomenclature}

%%%%%%%%%  BODY OF PAPER %%%%%%%%%%%%%%%%%%%%%%%%%%%%%%%%%
\section{Introduction}
Non-Newtonian complex liquids are important in applications ranging from red blood cell separation to interfacial phenomena in spray paints~\citep{kumar2023particle,kumar2025viscosity,kumar2026viscosity_jcis,baldygin2022effect}. Quantifying these particles or bubbles are very critical~\citep{nayak2026depth,madanan2021computational}. In future heat pipes research, shear-thinning behavior could reduces interfacial shear, thereby enhancing refrigerant transport~\citep{kumar2025experimental,kumar2024hydrostatic}. Yield-stress and shear-thinning fluids arise in a wide range of engineering and geophysical flows, including consumer products (e.g., Carbopol- and polymer-based formulations), slurry transport, and complex industrial processing. In fact, in pump design for pumping these kinds of high-viscosity liquids or slurry/bubbly flow~\cite{kumar2025bubble,javadi2026large}. A central feature of Herschel--Bulkley (HB) materials is the coexistence of yielded and weakly-yielded
(or unyielded) regions, which can produce plug-like cores and strongly non-uniform shear distributions. These
rheological effects directly influence momentum transport and the onset and organization of turbulence in
wall-bounded configurations \cite{reviewyusufi2025advances,pope2000turbulent}. 
Herschel--Bulkley rheology is also relevant in slurry transport, where the carrier fluid may exhibit both yield stress and shear-thinning behavior. These properties can lead to a plug-like core and a highly sheared wall region, which in turn affect the flow structure and pressure losses in pipes. This has been noted in studies of bentonite-based slurries, where the Herschel--Bulkley model provides a better description than simpler viscoplastic models when the shear-thinning response becomes important. For this reason, understanding transition in Herschel--Bulkley pipe and channel flows is also useful as a first step toward more realistic slurry-flow modeling \cite{wang2023numerical,yusufi2024rheology}.

For Newtonian pipe flow, transition is classically mediated by localized structures such as puffs and slugs,
and the associated intermittency has been documented extensively \cite{nishi2008puffs,draad1998transition,pope2000turbulent}.
When the viscosity becomes shear-dependent, and especially when a finite yield stress is present, the transition
scenario can change substantially: the flow may exhibit strong spatial intermittency, long-lived coherent
structures, and pronounced asymmetry in the cross-section even under nominally symmetric forcing
\cite{escudier2005transition,peixinho2005laminar,esmael2008yield,guzel2009observation}.
Experiments on yield-stress and shear-thinning pipe flows have reported robust transitional dynamics,
including intermittent turbulent episodes and geometry-dependent asymmetry, across a range of rheologies
and operating conditions \cite{guzel2009intermittency,bahrani2014intermittency,charles2024asymmetry}.
Related transitions have also been explored in other non-Newtonian settings such as fiber suspensions and
yield power-law flows in annuli, emphasizing the broader relevance of plug-dominated laminar states and their
breakdown \cite{nikbakht2014fibre,erge2015annuli}.

On the modeling side, Reynolds-number scaling and transition correlations for non-Newtonian pipe flows have a long history,
beginning with early generalized correlations and continuing through modern analyses and datasets
\cite{metzner1955flow,reviewyusufi2025advances}.
For the turbulent regime, DNS and high-fidelity simulations of shear-thinning pipe flows have clarified how
shear-dependent viscosity modifies near-wall dynamics and turbulence statistics \cite{rudman2004turbulent,singh2017influence}.
In channel flow, canonical Newtonian DNS benchmarks remain a primary reference point for turbulence statistics
and spectral content \cite{kim1987turbulence,pope2000turbulent}, while recent studies have started to address
generalized-Newtonian and yield-stress effects in turbulent channels \cite{karahan2023turbulent}.
However, despite these advances, a unified DNS picture that resolves the complete laminar--transitional--turbulent
sequence in HB fluids including plug formation, plug breakdown, and the emergence of wall turbulence and that
does so consistently across both pipe and planar/rectangular channel geometries remains limited.

The objective of the present work is to perform direct numerical simulations of HB yield-stress fluids in
pressure-driven pipe and rectangular channel configurations over a broad range of generalized Reynolds numbers.
We focus on resolving the transition dynamics and the accompanying changes in near-wall and core-region behavior,
with particular attention to plug-dominated states and their evolution. For the pipe, we validate key transition
metrics against experimental trends reported for Carbopol flows \cite{guzel2009observation}. For the channel,
we use a fully resolved three-dimensional domain and analyze statistical and spectral measures following common
wall-bounded turbulence \cite{kim1987turbulence,pope2000turbulent,prajapati2025laminar}.

\section{Methodology}
\label{sec:methodology}
\subsection{Rheology model}
\label{subsec:rheology}

We model the working fluid as a Herschel--Bulkley (HB) material characterized by yield stress $\tau_y$, consistency index $k$, and power-law index $n$. In simple shear, the HB constitutive relation is
\begin{equation}
\tau = \tau_y + k \dot{\gamma}^{\,n}, \qquad \text{for } \tau>\tau_y,
\label{eq:HB_tau}
\end{equation}
and the material remains unyielded when the applied stress is below the yield stress, so that $\dot{\gamma}=0$ in that regime \citep{karahan2023turbulent}.
Setting $\tau_y=0$ recovers the power-law model, while $(n=1,\tau_y=0)$ gives the Newtonian limit.

For three-dimensional incompressible generalized-Newtonian simulations, we write the viscous stress tensor in terms of the strain-rate tensor $S_{ij}$ as
\begin{equation}
\tau_{ij} = 2 \nu(\dot{\gamma})\, S_{ij},
\label{eq:tauij}
\end{equation}
where $\nu=\mu/\rho$ is the kinematic viscosity. The effective strain rate is defined from the second invariant of $S_{ij}$ as
\begin{equation}
\dot{\gamma} = \sqrt{2 S_{kl}S_{kl}}.
\label{eq:gammadot}
\end{equation}
For an HB fluid, the apparent kinematic viscosity is written as
\begin{equation}
\nu(\dot{\gamma}) = \frac{1}{\rho}\left(\frac{\tau_y}{\dot{\gamma}} + k \dot{\gamma}^{\,n-1}\right).
\label{eq:nu_HB}
\end{equation}
These expressions follow the standard generalized-Newtonian HB formulation used in prior DNS studies \citep{karahan2023turbulent,reviewyusufi2025advances}.

Equation~\eqref{eq:nu_HB} becomes numerically stiff at very small $\dot{\gamma}$, and directly resolving an ideal yielded/unyielded split is challenging. We therefore use a bi-viscosity regularization in which the viscosity is capped below a cut-off strain rate $\dot{\gamma}_0$:
\begin{equation}
\nu(\dot{\gamma}) = \min\left(\nu_0,\; \frac{1}{\rho}\left(\frac{\tau_y}{\dot{\gamma}} + k \dot{\gamma}^{\,n-1}\right)\right),
\label{eq:bivisc}
\end{equation}
with
\begin{equation}
\nu_0 = \frac{1}{\rho}\left(\frac{\tau_y}{\dot{\gamma}_0} + k \dot{\gamma}_0^{\,n-1}\right).
\label{eq:nu0}
\end{equation}
The cut-off $\dot{\gamma}_0$ is chosen small enough that it primarily affects nearly unyielded zones, while the yielded-region statistics remain unchanged within the accuracy of the simulations \citep{karahan2023turbulent,reviewyusufi2025advances}.

\subsection{Non-Newtonian Reynolds number}
\label{subsec:reynolds}

For HB fluids there is no single universally accepted Reynolds number because the viscosity varies with shear rate and the yield stress can produce unyielded regions \citep{reviewyusufi2025advances}. Two conventions are particularly common in the literature. The first is the Metzner--Reed (MR) approach, originally introduced to extend Newtonian friction-factor correlations to non-Newtonian fluids \citep{metzner1955flow}. The second is to define a generalized Reynolds number using a wall-viscosity scale, which is often convenient for wall-bounded transition and turbulence \citep{rudman2004turbulent,singh2017influence,karahan2023turbulent}.

\paragraph{Generalized Reynolds number based on wall viscosity.}
we define the generalized Reynolds number as
\begin{equation}
Re_G = \frac{\rho U L}{\mu_w},
\label{eq:ReG}
\end{equation}
where $U$ is the bulk velocity and $L$ is a characteristic length scale (pipe diameter $D$ for pipe flow, and channel half-height $h$ for channel flow). For HB fluids, the mean wall dynamic viscosity can be expressed in terms of the mean wall shear stress $\tau_w$ as
\begin{equation}
\mu_w = k^{1/n}\,\frac{\tau_w}{(\tau_w-\tau_y)^{1/n}}.
\label{eq:muw}
\end{equation}
For a planar channel driven by a constant mean pressure gradient, $\tau_w = -h\, dP/dx$, while for a pipe $\tau_w$ can be obtained from the imposed pressure gradient in the conventional way \citep{karahan2023turbulent,reviewyusufi2025advances}. While engineering approaches such as the Chilton--Stainsby framework are useful for transport and pressure-loss estimation, the present study is aimed at resolving the physics of laminar-to-turbulent transition in Herschel--Bulkley flows through DNS, rather than developing a new correlation for design purposes.

\paragraph{Metzner--Reed Reynolds number.}
The Metzner--Reed idea is widely used for shear-dependent fluids and is commonly reported in turbulent pipe-flow studies of shear-thinning fluids \citep{metzner1955flow,rudman2004turbulent,singh2017influence}. For power-law fluids, a frequently used closed form is
\begin{equation}
Re_{MR}=\frac{8\,\rho\,U^{\,2-n}\,D^{\,n}}{k\left(6+\frac{2}{n}\right)^{n}},
\label{eq:ReMR_PL}
\end{equation}
which reduces to the Newtonian Reynolds number when $n=1$.

For HB fluids, using a power-law $Re_{MR}$ without explicitly accounting for $\tau_y$ does not collapse the laminar friction-factor relation across different yield-stress conditions. For this reason, a yield-stress-corrected Metzner--Reed Reynolds number for steady laminar planar channel flow can be written as \citep{karahan2023turbulent}
\begin{equation}
Re_{MR,HB} =
\frac{\rho U h}{\mu_w}
\left(\frac{1+2n}{3n}\right)
\left(\frac{1}{1-a\chi-b\chi^2}\right),
\label{eq:ReMRHB}
\end{equation}
where
\begin{equation}
a=\frac{1}{n+1}, \qquad
b=\frac{n}{n+1}, \qquad
\chi=\frac{\tau_y}{\tau_w}.
\label{eq:abchi}
\end{equation}
This construction ensures that the laminar friction relation collapses in a consistent way for HB channel flows \citep{karahan2023turbulent}. We therefore report both $Re_G$ and MR-based Reynolds numbers to enable direct comparison with the range of conventions used in experiments and DNS \citep{peixinho2005laminar,guzel2009observation,reviewyusufi2025advances}.

\paragraph{Friction Reynolds number.}
For inner scaling and near-wall normalization, we also use a friction Reynolds number based on the same wall-viscosity scale,
\begin{equation}
Re_\tau = \frac{\rho u_\tau h}{\mu_w},
\qquad
u_\tau = \sqrt{\frac{\tau_w}{\rho}},
\label{eq:Retau}
\end{equation}

\subsection{Governing equations}
\label{subsec:gov_eq}

We solve the incompressible momentum equation in a form that is convenient for shear-dependent viscosity. For notational simplicity, the pressure, stress, and forcing are written in kinematic form (divided by the constant density), while we continue to refer to them as pressure, stress, and body force. Following this convention, the governing equation is
\begin{equation}
\frac{\partial \mathbf{u}}{\partial t} + \mathbf{u}\cdot\nabla\mathbf{u}
= -\nabla p + \nabla\cdot\boldsymbol{\tau} + \mathbf{g},
\label{eq:gov_21}
\end{equation}

The stress tensor is modeled using the generalized-Newtonian (GN) assumption,
\begin{equation}
\boldsymbol{\tau} = 2\nu(\dot{\gamma})\,\mathbf{s},
\label{eq:gov_22}
\end{equation}
where \(\mathbf{s}\) is the instantaneous strain-rate tensor (defined in Section~3.1) and \(\nu(\dot{\gamma})\) is the apparent kinematic viscosity. In the present work, \(\nu(\dot{\gamma})\) is obtained from the regularized Herschel--Bulkley model described in Section~3.1, rather than a power-law model. \citep{singh2017influence,karahan2023turbulent}
% Support for GN stress structure: :contentReference[oaicite:2]{index=2}

For numerical robustness, the nonlinear term is implemented in a skew-symmetric form, so that the discrete system has improved stability properties. The implemented momentum equation is
\begin{equation}
\frac{\partial \mathbf{u}}{\partial t}
+\frac{1}{2}\left(\mathbf{u}\cdot\nabla\mathbf{u}+\nabla\cdot(\mathbf{u}\mathbf{u})\right)
= -\nabla p + \nabla\cdot\left(2\nu\mathbf{s}\right) + \mathbf{g}.
\label{eq:gov_23}
\end{equation}
The forcing \(\mathbf{g}\) acts only in the streamwise direction and is set to achieve the desired bulk flow rate, while allowing the pressure field to remain periodic. \citep{singh2017influence}
% Support: :contentReference[oaicite:3]{index=3}

When the time integration scheme benefits from treating a constant-viscosity Laplacian implicitly, the viscosity can be split into a spatially uniform reference value \(\nu_{\mathrm{ref}}\) plus a variable remainder. This yields
\begin{equation}
\frac{\partial \mathbf{u}}{\partial t}
+\frac{1}{2}\left(\mathbf{u}\cdot\nabla\mathbf{u}+\nabla\cdot(\mathbf{u}\mathbf{u})\right)
= -\nabla p
+ \nu_{\mathrm{ref}}\nabla^2\mathbf{u}
+ 2\nabla\cdot\left\{\left(\nu-\nu_{\mathrm{ref}}\right)\mathbf{s}\right\}
+ \mathbf{g}.
\label{eq:gov_24}
\end{equation}
In this form, the term \(\nu_{\mathrm{ref}}\nabla^2\mathbf{u}\) can be advanced implicitly, while the remaining viscosity-dependent contribution is handled explicitly together with the other nonlinear terms, enabling stable time steps close to the CFL limit. \citep{singh2017influence}
% Support: :contentReference[oaicite:4]{index=4}

\section{Domain discretization}
\label{sec:domain_discretization}

\subsection{Computational domains and mesh}
\label{subsec:domains_mesh}

\begin{figure}[t]
  \centering
  \includegraphics[width=0.40\textwidth]{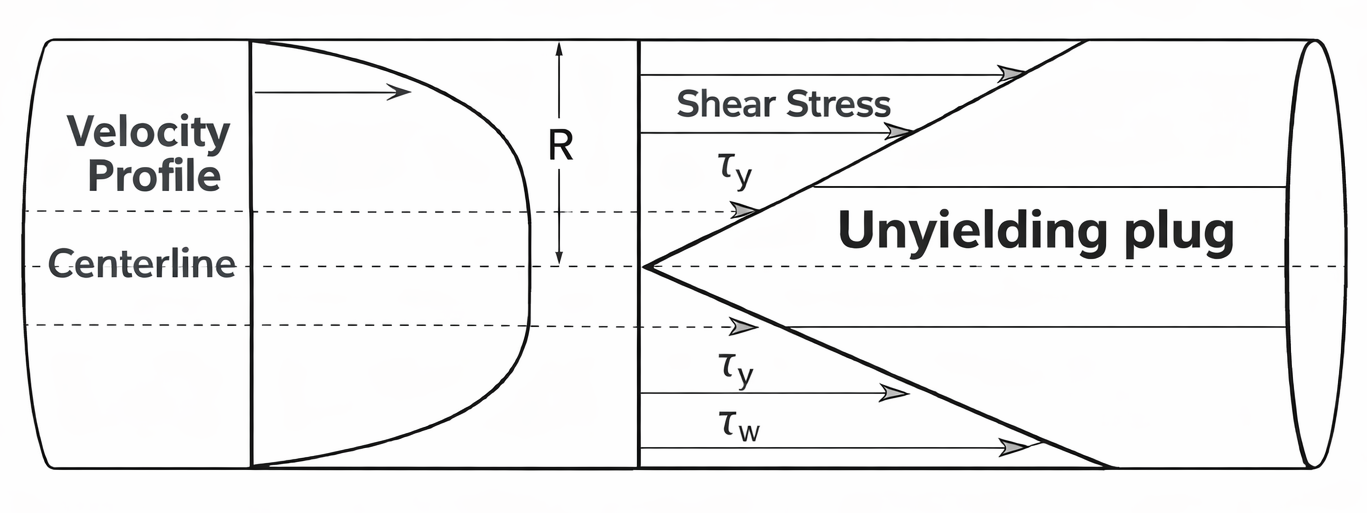}
  \caption{Pipe flow geometry and coordinate system.}
  \label{fig:pipe_geometry}
\end{figure}

\paragraph{Pipe flow.}
The pipe domain has a circular cross-section with radius \(R = 25.4~\mathrm{mm}\) (diameter \(D=2R\)) and a streamwise length
\(L = 4\pi D\), as illustrated in Fig.~\ref{fig:pipe_geometry}. The simulations use a structured O-grid topology generated with
OpenFOAM's \texttt{blockMesh}. The O-grid provides smooth cell transitions and avoids excessively skewed cells near the wall.
The radial spacing is clustered toward the wall using a geometric progression to better resolve the near-wall gradients and the
interface between yielded and weakly sheared regions. The resulting mesh contains approximately \(4\times 10^{6}\) control volumes.

\begin{figure}[t]
  \centering
  \includegraphics[width=0.40\textwidth]{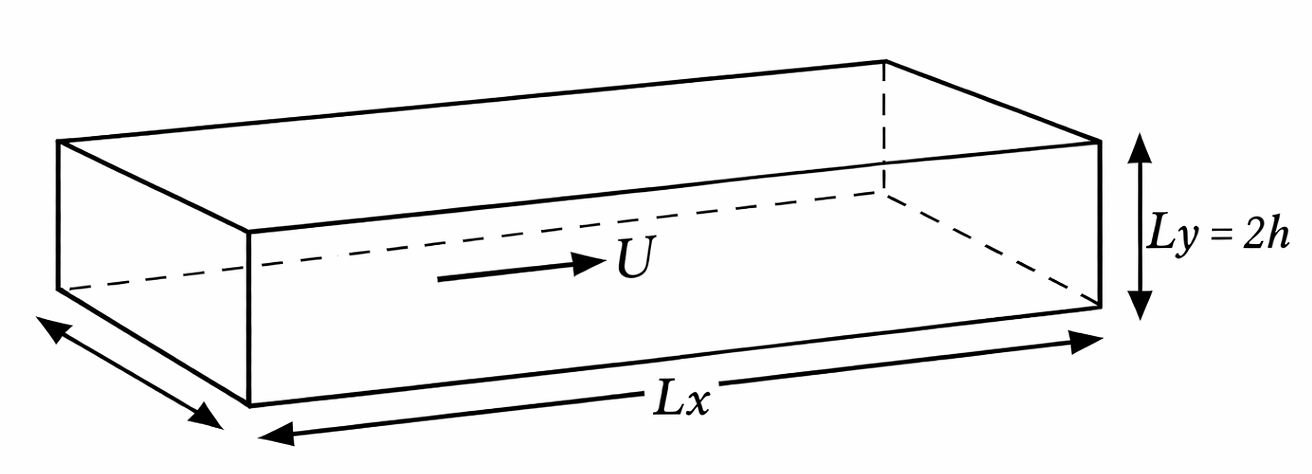}
  \caption{Channel flow geometry and coordinate system.}
  \label{fig:channel_geometry}
\end{figure}

\paragraph{Channel flow.}
The channel domain is a rectangular box with streamwise length \(L_x = 0.2~\mathrm{m}\), as shown in Fig.~\ref{fig:channel_geometry}.
The streamwise, wall-normal, and spanwise directions are denoted by \(x\), \(y\), and \(z\), respectively, with total channel height
\(L_y = 2h\). The channel simulations use a structured mesh of \(128 \times 129 \times 128\) points in \((x,y,z)\), corresponding to
approximately \(4.2\times 10^{6}\) cells.

A mesh-independence assessment was used to select the final grid. For both pipe and channel cases, the near-wall resolution satisfies
\(y^+<1\) based on the friction velocity and the first-cell center location.

% ===================== Add this inside Domain Discretization (same subsection) =====================

\section{Results \& discussion}
\subsection{Domain-size sufficiency via two-point correlations.}
For a periodic domain, domain-size sufficiency is assessed by verifying that the two-point correlations of velocity fluctuations converge to near zero at large separations in the periodic directions, indicating that the dominant structures are well contained within the computational domain. Two-point correlations of velocity fluctuations provide a direct and commonly used diagnostic for this purpose \citep{pope2000turbulent,kim1987turbulence}. In the channel, we therefore compute correlations in the streamwise direction at two representative wall-normal locations: a near-wall position and a position near the channel center. This is important because near-wall motions and outer-layer motions can exhibit different coherence lengths and characteristic spacings \citep{pope2000turbulent}.

\begin{figure}[t]
  \centering
  \begin{subfigure}[t]{0.4\textwidth}
    \centering
    \includegraphics[width=\textwidth]{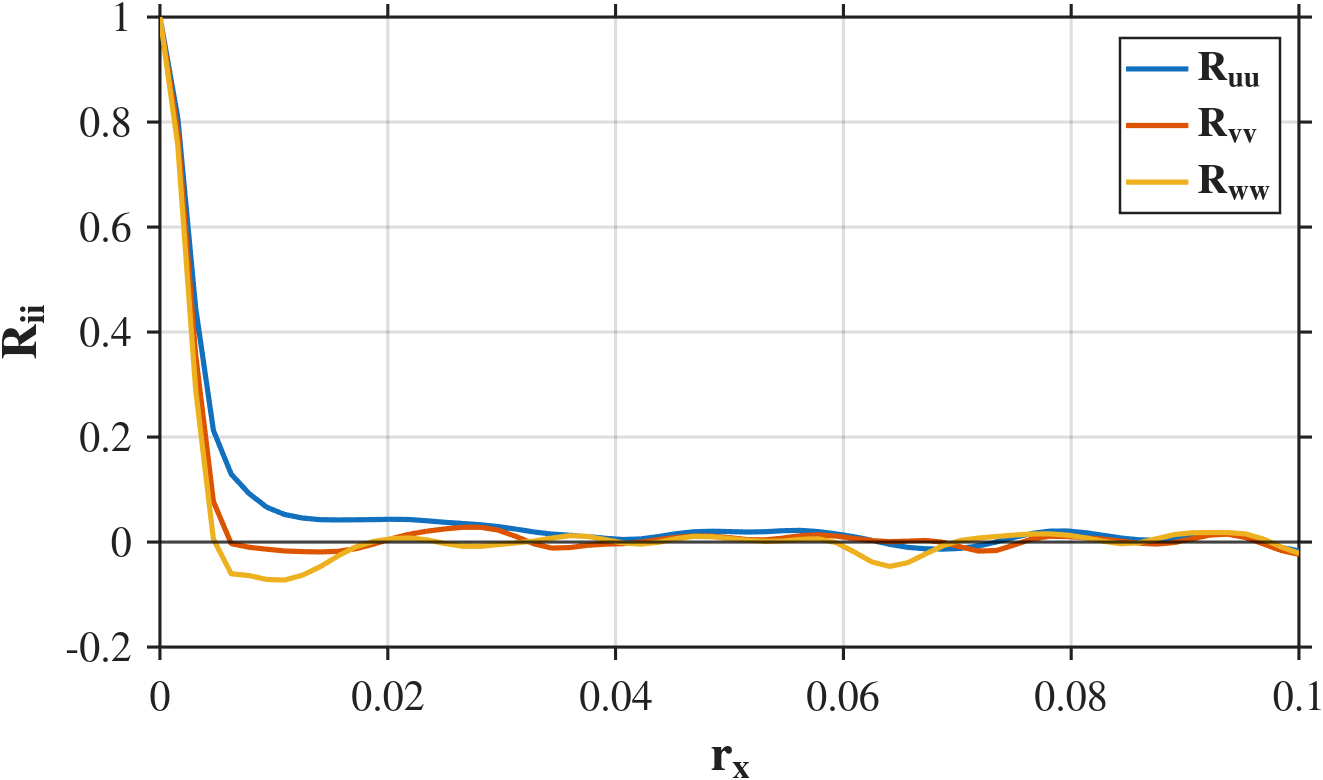}
    \caption{Near-wall location.}
  \end{subfigure}
  \hfill
  \begin{subfigure}[t]{0.4\textwidth}
    \centering
    \includegraphics[width=\textwidth]{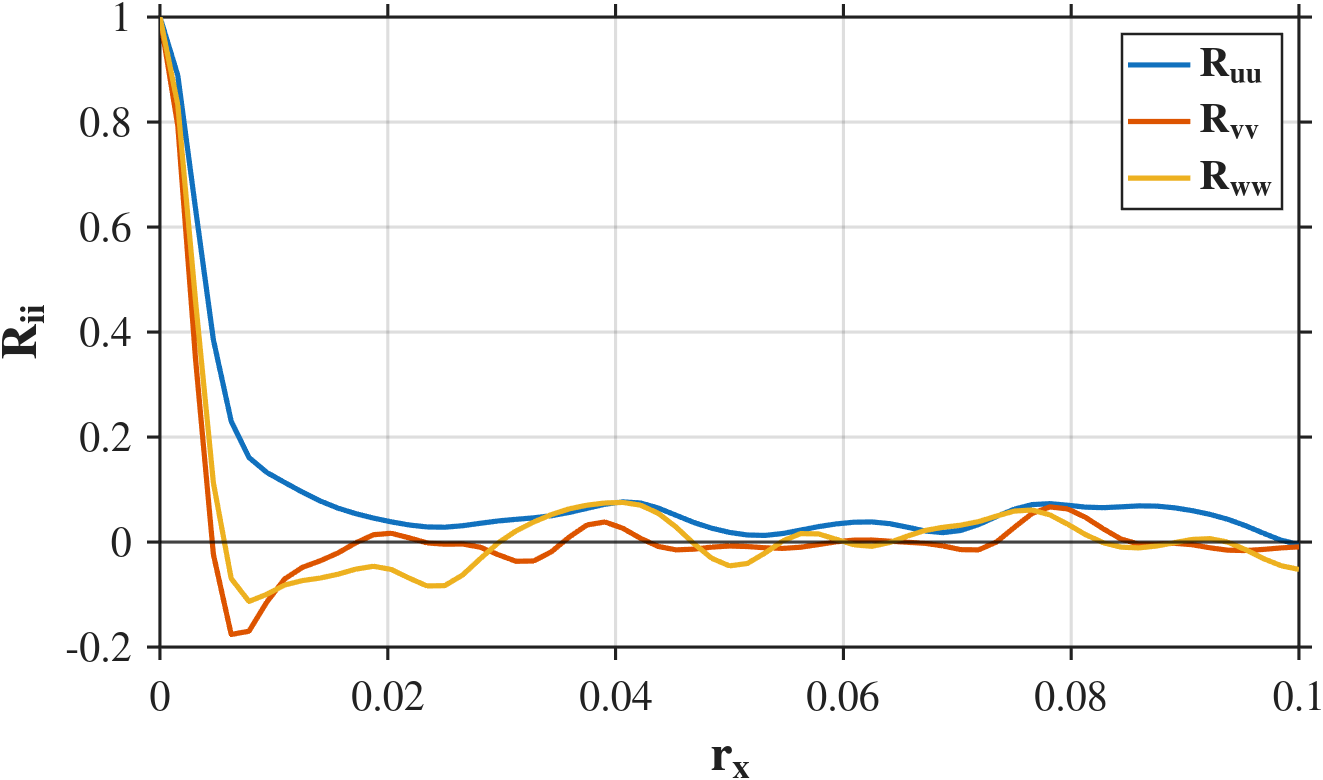}
    \caption{Near-center location.}
  \end{subfigure}
  \caption{Streamwise two-point correlations in the channel for the velocity components \(u\), \(v\), and \(w\): \(R_{uu}(r_x;y)\), \(R_{vv}(r_x;y)\), and \(R_{ww}(r_x;y)\).}
  \label{fig:tpc_streamwise}
\end{figure}

For the channel, where \(x\) is streamwise, \(y\) is wall-normal, and \(z\) is spanwise, we define the streamwise two-point correlations of the velocity fluctuations as
\begin{equation}
R_{ii}(r_x; y) \;=\;
\frac{\left\langle u_i'(x,y,z,t)\,u_i'(x+r_x,y,z,t)\right\rangle_{z,t}}
{\left\langle u_i'^2(x,y,z,t)\right\rangle_{z,t}},
\qquad i\in\{u,v,w\},
\label{eq:Rii_streamwise}
\end{equation}
where \(u_i'=\{u',v',w'\}\) are the fluctuating velocity components obtained after subtracting the time-averaged mean, and \(\langle\cdot\rangle\) denotes averaging over the directions and time indicated by the subscripts. By construction, \(R_{ii}(0;y)=1\). The decay of \(R_{ii}\) with separation provides a quantitative measure of the coherence length of the corresponding velocity component and a practical check for periodic-domain sufficiency \citep{pope2000turbulent,kim1987turbulence}. Figure~\ref{fig:tpc_streamwise} shows the streamwise correlations \(R_{uu}(r_x;y)\), \(R_{vv}(r_x;y)\), and \(R_{ww}(r_x;y)\) at the near-wall location (Fig.~\ref{fig:tpc_streamwise}a) and near the channel center (Fig.~\ref{fig:tpc_streamwise}b).

\subsection{One-dimensional energy spectra}
\label{subsec:spectra}

One-dimensional energy spectra are computed as a standard DNS diagnostic to quantify how turbulent kinetic energy is distributed across length scales in the homogeneous directions and to assess whether the selected grid resolution and domain size provide an adequate representation of the resolved range \citep{kim1987turbulence}. For the channel, we evaluate spectra in the streamwise direction.

Let $u_i'(\mathbf{x},t)=u_i(\mathbf{x},t)-\overline{u_i}(y)$ denote the fluctuating velocity component, where the overbar denotes averaging over time and the homogeneous directions at fixed wall-normal coordinate $y$. For a fixed $y$, we compute a Fourier transform in the streamwise direction,
\begin{equation}
\widehat{u_i'}(k_x,y,z,t)=\int_{0}^{L_x} u_i'(x,y,z,t)\,e^{-\,\mathrm{i}k_x x}\,\mathrm{d}x,
\label{eq:fourier_x}
\end{equation}
and form the one-dimensional streamwise energy spectrum of component $i$ by averaging in time and integrating over the spanwise wavenumber,
\begin{equation}
E_{ii}(k_x,y)=\left\langle \int |\widehat{u_i'}(k_x,k_z,y,t)|^2 \,\mathrm{d}k_z \right\rangle_t .
\label{eq:spectrum_kx}
\end{equation}

\begin{figure}[t]
  \centering
  \includegraphics[width=0.48\textwidth]{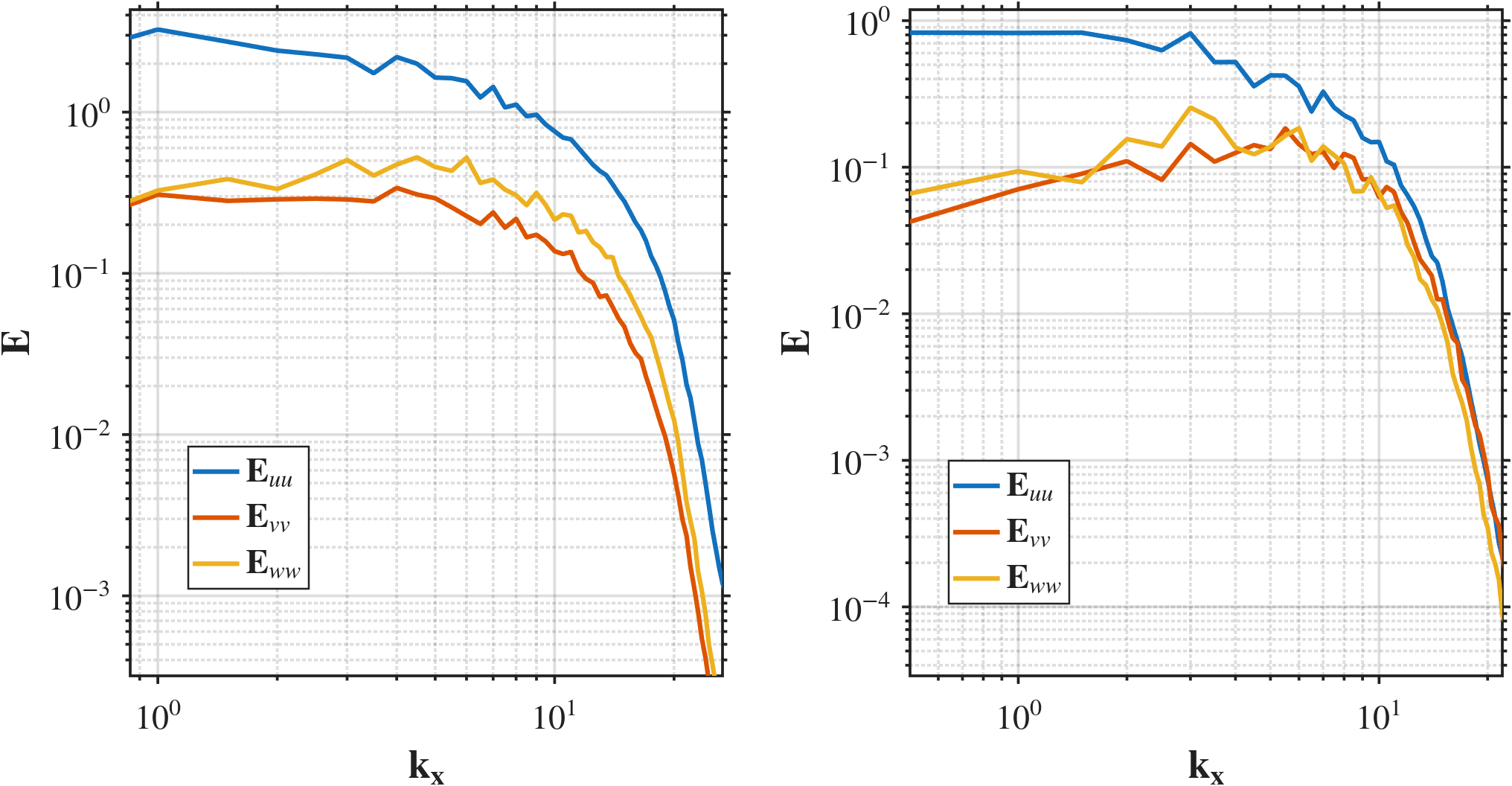}
  \caption{One-dimensional streamwise energy spectra in the channel at a near-wall location and at the channel center: $E_{uu}(k_x)$, $E_{vv}(k_x)$, and $E_{ww}(k_x)$.}
  \label{fig:spectra_kx}
\end{figure}

In the present channel simulations, we report the streamwise spectra $E_{uu}(k_x)$, $E_{vv}(k_x)$, and $E_{ww}(k_x)$ at a near-wall plane and at the channel center in Fig.~\ref{fig:spectra_kx}. These spectra are used as scale-by-scale diagnostics of the resolved turbulent motions in the streamwise direction, consistent with established channel-flow DNS studies \citep{kim1987turbulence}.

\section{Validation against experiments for pipe flow}
\label{sec:validation}

To validate the numerical framework, we compare our pipe-flow simulations against the laser doppler velocimetry (LDV) measurements reported by Guzel et al. \citep{guzel2009observation} for 0.1\% Carbopol solutions in a circular pipe. The experimental dataset spans laminar, transitional, and turbulent regimes, which makes it well suited for checking whether the present solver, rheology implementation, and numerical resolution reproduce the observed onset and development of turbulence in a yield-stress, shear-thinning fluid.

In our validation cases, we match the Herschel--Bulkley parameters \((\tau_y, k, n)\) and consider the same bulk-velocity range as the experiments. Table~\ref{tab:guzel_conditions} summarizes the operating conditions and rheological parameters used for comparison. 

\begin{table}[t]
\centering
\caption{Flow conditions and Herschel--Bulkley parameters for 0.1\% Carbopol pipe flow, adapted from Guzel et al. \citep{guzel2009observation}.}
\label{tab:guzel_conditions}
\begin{tabular}{c c c c c c}
\hline
$U$ (m/s) & $\dot{\gamma}$ (s$^{-1}$) & $\tau_y$ (Pa) & $k$ (Pa$\cdot$s$^n$) & $n$ & $Re_G$ \\
\hline
1.20760 & 1--220    & 1.4 & 1.59 & 0.43 & 378 \\
2.04610 & 5--414    & 1.3 & 1.20 & 0.48 & 937 \\
2.32180 & 5--472    & 1.2 & 0.92 & 0.53 & 1160 \\
3.11460 & 5--667    & 1.0 & 0.65 & 0.60 & 1735 \\
3.90050 & 5--1261   & 0.6 & 0.35 & 0.65 & 2920 \\
4.39670 & 5--1559   & 0.4 & 0.20 & 0.70 & 4488 \\
\hline
\end{tabular}
\end{table}

The validation focuses on comparisons that are directly accessible in both experiments and simulations. Specifically, we compare (i) the   streamwise velocity profile in wall units and in outer scaling, (ii) turbulence intensity trends with increasing $Re_G$, and (iii) qualitative changes in profile shape associated with the breakdown of the plug-like core as the flow transitions. Agreement across these metrics provides confidence that the present numerical setup reproduces the experimentally observed regime progression and the associated changes in near-wall and core-region dynamics for Herschel--Bulkley pipe flow \citep{guzel2009observation}.
\subsection{Validation using turbulence intensity trends}
\label{subsec:validation_TI}

To validate the numerical framework against experiments, we compare the radial variation of turbulence intensity as a function of the generalized Reynolds number, using the 0.1\% Carbopol pipe-flow measurements of Guzel et al.\citep{guzel2009observation}. Turbulence intensity is used here as a compact measure of the strength of velocity fluctuations relative to the mean flow. At a given radial location, we define
\begin{equation}
I(r) = 100\,\frac{u'_{\mathrm{rms}}(r)}{U_b},
\label{eq:TI_def}
\end{equation}
where \(u'_{\mathrm{rms}}(r)\) is the root-mean-square of the streamwise velocity fluctuations at radius \(r\), and \(U_b\) is the bulk (cross-section averaged) velocity. This definition is consistent with the experimental reporting of turbulence intensity trends and enables a direct comparison between DNS and LDV data.

\begin{figure}[t]
  \centering
  \includegraphics[width=0.4\textwidth]{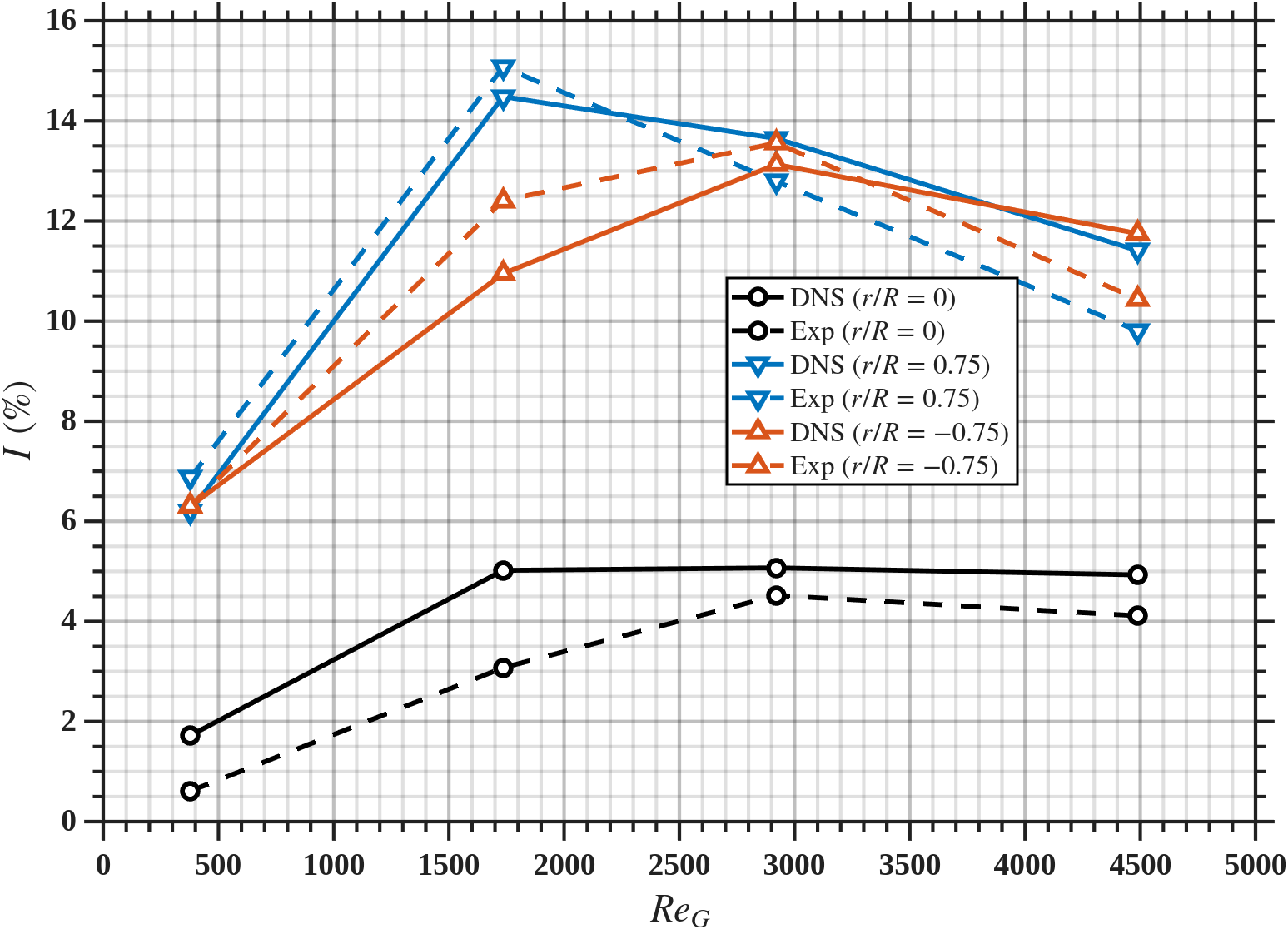}
  \caption{Turbulence intensity $I$ (\%) as a function of generalized Reynolds number $Re_G$ in Herschel--Bulkley pipe flow. Solid lines with markers show the present DNS, and dashed lines show the experimental data. Results are reported at $r/R=0$, $r/R=0.75$, and $r/R=-0.75$.}
  \label{fig:TI_ReG_pipe_validation}
\end{figure}

Figure~\ref{fig:TI_ReG_pipe_validation} shows the measured and computed turbulence intensity at three locations, \(r/R=0\) and \(r/R=\pm 0.75\). At the lowest Reynolds number considered (\(Re_G \approx 378\)), the flow is weakly disturbed and remains largely plug-dominated. Yield-stress effects suppress deformation in the core and reduce velocity fluctuations, leading to low turbulence intensity at the centerline. The near-wall locations already exhibit larger values than the centerline because the highest shear occurs in the wall region, where the material is more likely to be yielded and susceptible to unsteady motions.

As \(Re_G\) increases into the transitional regime (\(Re_G \approx 1735\) to \(2920\)), inertial effects become strong enough to overcome the yield-stress constraint over a larger portion of the cross-section. This promotes the growth of localized instabilities, strengthens near-wall streak-like motions, and accelerates the breakdown of the plug boundary. As a result, turbulence intensity rises sharply, with the strongest fluctuations occurring away from the centerline at \(r/R=\pm 0.75\). Both the experiments and the simulations display this behaviour, including the non-monotonic character of the off-center intensity across the transitional window.

At higher Reynolds number (\(Re_G \approx 4488\)), the flow becomes more broadly turbulent and the intensity decreases slightly at \(r/R=\pm 0.75\) compared with its transitional peak. This reduction can occur because the mean flow increases with \(Re_G\) while the relative fluctuation level does not grow proportionally, and because the plug resistance becomes less dominant as yielding extends further into the core. Overall, the close agreement between DNS and experiment in terms of (i) the strong radial dependence, (ii) the sharp rise through transition, and (iii) the peak-and-decline trend at the off-center locations supports the fidelity of the present numerical setup for Herschel--Bulkley pipe flow. 
A small asymmetry is also observed in the turbulence intensity trends. The $I$--$Re_G$ curves at $r/R=\pm 0.75$ do not coincide perfectly, particularly through the transitional range, consistent with prior experimental reports of asymmetric or biased states in transitional yield-stress and shear-thinning pipe flows (e.g., \citep{guzel2009observation,peixinho2005laminar}).
\subsection{Vortical structures via Q-criterion}
\label{subsec:qcriterion}

To visualize coherent vortical motions in three-dimensional pipe flow, we use the Q-criterion, which identifies regions where local rotation dominates over strain \citep{pope2000turbulent}. For an incompressible flow, Q is defined from the velocity-gradient tensor $\nabla\mathbf{u}$ as
\begin{equation}
Q=\frac{1}{2}\left(\|\boldsymbol{\Omega}\|^{2}-\|\mathbf{S}\|^{2}\right),
\label{eq:Qcriterion}
\end{equation}
where $\mathbf{S}=\tfrac{1}{2}\left(\nabla\mathbf{u}+(\nabla\mathbf{u})^{T}\right)$ is the strain-rate tensor and
$\boldsymbol{\Omega}=\tfrac{1}{2}\left(\nabla\mathbf{u}-(\nabla\mathbf{u})^{T}\right)$ is the rotation-rate tensor. Positive values of $Q$ indicate rotation-dominated regions, and iso-surfaces of $Q$ provide a convenient way to highlight vortical structures in transitional and turbulent flow fields.

\begin{figure}[t]
    \centering
    \begin{minipage}{0.42\textwidth}
        \centering
        \includegraphics[width=\linewidth]{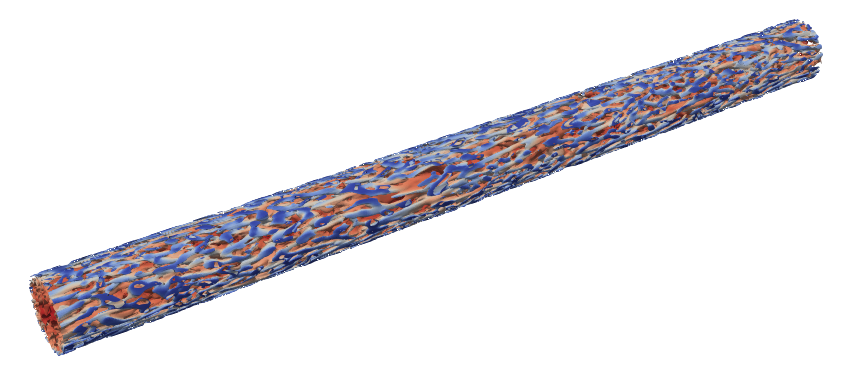}
    \end{minipage}
    \vskip0.3em
    \begin{minipage}{0.40\textwidth}
        \centering
        \includegraphics[width=\linewidth]{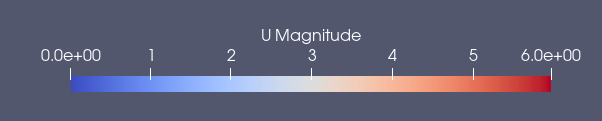}
    \end{minipage}
    \caption{Q-criterion iso-surfaces for $Re_G=4488$ with the corresponding velocity-magnitude legend. The field shows dense small-scale vortical activity consistent with turbulent dynamics.}
    \label{fig:q-turbulent}
\end{figure}

\begin{figure}[t]
    \centering
    \begin{minipage}{0.42\textwidth}
        \centering
        \includegraphics[width=\linewidth]{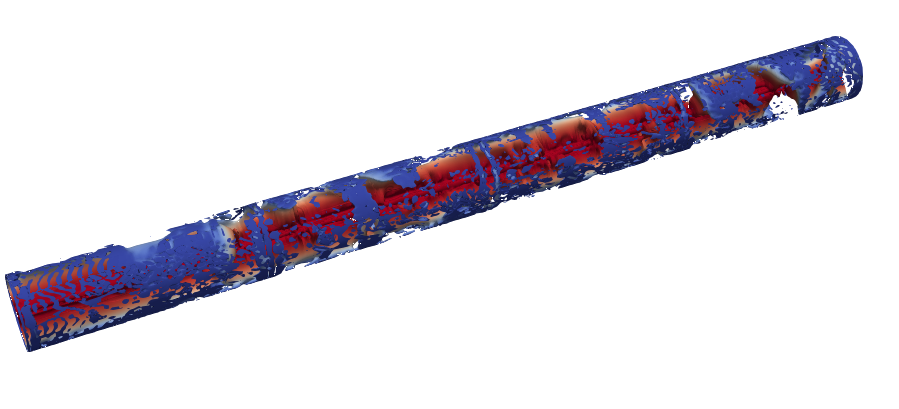}
    \end{minipage}
    \vskip0.3em
    \begin{minipage}{0.40\textwidth}
        \centering
        \includegraphics[width=\linewidth]{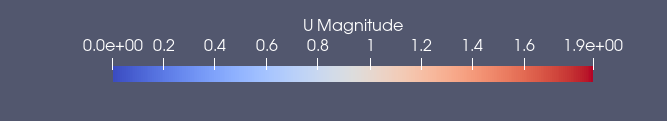}
    \end{minipage}
    \caption{Q-criterion iso-surfaces for $Re_G=378$ with the corresponding velocity-magnitude legend. Vortical structures are comparatively sparse and more coherent, consistent with a laminar or weakly disturbed regime.}
    \label{fig:q-laminar}
\end{figure}

Figures~\ref{fig:q-turbulent} and~\ref{fig:q-laminar} show representative $Q$-criterion iso-surfaces for two cases, $Re_G=4488$ and $Re_G=378$, together with the corresponding velocity-magnitude legends. At $Re_G=4488$, the iso-surfaces reveal abundant small-scale vortical activity throughout the pipe, consistent with a fully developed turbulent state. In contrast, at $Re_G=378$ the flow exhibits comparatively sparse vortical content, with structures that are larger and more organized. This behavior is consistent with a laminar or weakly disturbed regime in which coherent vortical motion is largely confined to the sheared region near the wall, while the core remains weakly sheared due to yield-stress effects.

Overall, the Q-criterion visualizations provide a qualitative view of how coherent flow structures evolve across regimes in Herschel--Bulkley pipe flow, complementing the statistical measures used to classify laminar, transitional, and turbulent behavior.

\section{Channel flow}
\begin{figure}[t]
  \centering
  \includegraphics[width=0.95\linewidth]{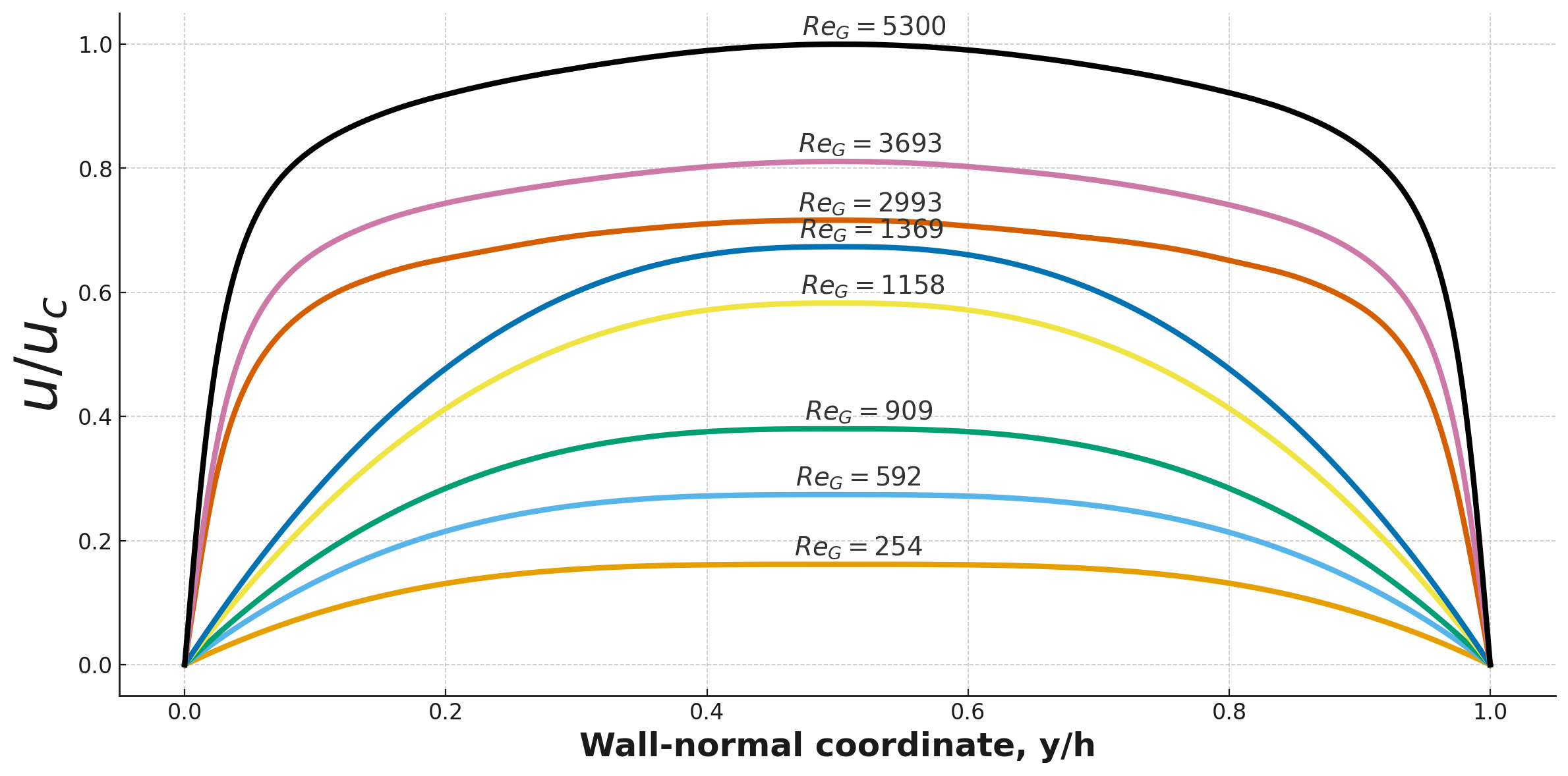}
  \caption{Normalized, time-averaged streamwise velocity profiles ($u/u_c$) in the channel at
  increasing generalized Reynolds number $Re_G$ for the 0.1\% Carbopol Herschel--Bulkley fluid.}
  \label{fig:Umean_channel_ReG}
\end{figure}

Table~\ref{tab:channel_flow_parameters} summarizes the operating conditions and nondimensional groups for the rectangular channel simulations of the 0.1\% Carbopol solution modeled as a Herschel--Bulkley fluid. Each case is defined by the imposed mean pressure gradient $\partial p/\partial x$ and the resulting bulk velocity $U$. For completeness, we also report the wall-based viscosity $\mu_w$, friction velocity $u_\tau$, and friction Reynolds number $Re_\tau$, along with the Herschel--Bulkley parameters $(\tau_y,\kappa,n)$ and the generalized Reynolds number $Re_G$ used to index the transition progression across cases.
% -------------------- Channel-flow cases: 0.1% Carbopol (HB) --------------------
\begin{table*}[t]
\centering
\caption{Channel-flow operating and Herschel--Bulkley parameters for the 0.1\% Carbopol cases used in the DNS. The fluid is modeled as Herschel--Bulkley and thixotropic effects are neglected.}
\label{tab:channel_flow_parameters}
\setlength{\tabcolsep}{6pt}
\renewcommand{\arraystretch}{1.15}
\resizebox{0.75\linewidth}{!}{%
\begin{tabular}{ccccccccc}
\hline
$U$ (m/s) & $\partial p/\partial x$ (Pa/m) & $\mu_w$ (Pa$\cdot$s) & $u_\tau$ (m/s) & $Re_\tau$ & $\tau_y$ (Pa) & $\kappa$ (Pa$\cdot$s$^{n}$) & $n$ & $Re_G$ \\
\hline
4.397 & 11.80 & 0.01807 & 0.23646 & 62.31  & 0.40 & 0.20 & 0.70 & 1158.69 \\
5.000 & 12.82 & 0.01739 & 0.24709 & 67.68  & 0.40 & 0.20 & 0.70 & 1369.51 \\
6.000 & 29.07 & 0.01218 & 0.37208 & 145.51 & 0.40 & 0.20 & 0.70 & 2346.45 \\
7.000 & 35.75 & 0.01000 & 0.40000 & 176.47 & 0.40 & 0.20 & 0.70 & 2993.59 \\
10.00 & 58.80 & 0.008986& 0.52918 & 280.46 & 0.40 & 0.20 & 0.70 & 5300.00 \\
\hline
\end{tabular}%
}
\end{table*}
% --- Mean velocity profiles (channel): text for your paper ---
Figure~\ref{fig:Umean_channel_ReG} shows the normalized, time-averaged streamwise velocity
profiles ($u/u_c$) for the 0.1\% Carbopol Herschel--Bulkley channel flow at increasing generalized
Reynolds number $Re_G$. At the lowest $Re_G$, the profile is strongly blunted, consistent with a
plug-dominated core where the shear rate is weak and most of the deformation is confined near the
walls. With increasing $Re_G$, the plug-like region progressively shrinks and the profile becomes
fuller across the channel, indicating that a larger fraction of the cross-section participates in the
sheared motion. At the highest $Re_G$, the mean profile approaches the familiar turbulent-channel
shape, with a fuller core and steeper near-wall gradients, reflecting the transition from plug-dominated
laminar flow to wall-dominated turbulence.

% -------------------------------------------------------------------------------

\section{Conclusion}
We performed direct numerical simulations (DNS) of Herschel--Bulkley (HB) yield-stress fluids in both pipe and rectangular channel configurations and resolved the complete progression from laminar flow to fully developed turbulence over a wide range of generalized Reynolds numbers. The simulations capture the yield-stress-driven plug, its gradual erosion with increasing inertia, and the emergence of near-wall turbulent activity during transition.

A key outcome is that transition is controlled by the local competition between turbulent momentum transport and the yield stress: sustained turbulence develops only when Reynolds stresses locally exceed the yield stress, allowing the plug-like core to deform and break down. For pipe flow, the DNS supports a clear regime classification:
\begin{itemize}
  \item $Re_G < 1735$: laminar regime with a strong plug and negligible turbulence,
  \item $1735 < Re_G < 2920$: transitional regime with intermittent/decaying turbulence,
  \item $Re_G > 2920$: fully turbulent regime where the plug breaks down and turbulence becomes established.
\end{itemize}

The computed turbulence-intensity trends are consistent with available experimental measurements for Carbopol pipe flow reported by G{\"u}zel et al.\ (2009), providing validation of the numerical framework. Finally, the pipe-flow results show evidence of asymmetry in the transitional range, consistent with observations reported in the literature for yield-stress and shear-thinning pipe flows.

%%%%% Acknowledgments %%%%%%%%%%%%%%%%%%%%%%%%%%%
\section*{Acknowledgments}
This material is based upon work supported by the U.S. Department of Energy’s Office of Energy Efficiency and Renewable Energy (EERE) under the Advanced Manufacturing Office, Award Number DE-EE0009396. Partial financial support was also provided by the Renewable Byproducts Institute Graduate Fellowship. The views expressed herein do not necessarily represent the views of the U.S. Department of Energy or the United States Government. In disclosure, the part of this work was presented at APS-DFD~\citep{prajapati2025laminar,suchandra2025surfactant,aps_kumar2025bubble_theory_dns,aps_kumar2024_exp}.

%%%  REFERENCES  %%%%%%%%%%%%%%%%%%%%%%%%%%%%%%%%
%%
%% Put your references into your .bib file in the usual way. Run latex once, bibtex once, then latex twice.
%% The asmeconf.bst style allows @inproceedings and @proceedings to include: 
%%		venue = {Location of Conference}, 
%%		eventdate = {Month, days},

\bibliographystyle{asmeconf}  %% .bst file following ASME conference format. Do not change.
\bibliography{asmeconf-sample}%% <=== change this to the name of your bib file

\end{document}